\documentclass[aps,prb,twocolumn,showpacs,superscriptaddress]{revtex4}

\usepackage{graphicx}
\usepackage{amsmath}

\begin{document}

\title{  Anisotropic magnetism and electronic properties of the kagome metal SmV$_6$Sn$_6$}

\author{Xing Huang}
\author{Zhiqiang Cui}
\author{Chaoxin Huang}
\author{Mengwu Huo}
\author{Hui Liu}
\author{Jingyuan Li}
\author{Feixiang Liang}
\author{Lan Chen}
\author{Hualei Sun}
\author{Bing Shen}
\author{Yunwei Zhang}
\author{Meng Wang}
\email{wangmeng5@mail.sysu.edu.cn}
\affiliation{Center for Neutron Science and Technology, Guangdong Provincial Key Laboratory of Magnetoelectric Physics and Devices, School of Physics, Sun Yat-Sen University, Guangzhou, 510275, China }

\begin{abstract}
Kagome magnets are expected to feature emergent properties due to the interplays among geometry, magnetism, electronic correlation, and band topology. The magnetism and topological electronic states can be tuned via the rare earth engineering in $R$V$_6$Sn$_6$ kagome metals, where $R$ is a rare earth element. Herein, we present the synthesis and characterization of SmV$_6$Sn$_6$, a metal with two-dimensional kagome nets of vanadium and frustrated triangular Sm lattice. Partial of the Sm atoms are shifted from the normal $R$ positions by $c$/2 along the $c$ axis. Magnetic measurements reveal obvious anisotropy, where the easy magnetic axis is within the $ab$ plane. Electronic transports show multiband behaviors below 200 K. Density functional theory calculations find that the electronic structure of SmV$_6$Sn$_6$ hosts flat bands, Dirac cone, and saddle point arising from the V-3$d$ electrons near the Fermi level. No evidence for the existence of charge density wave or magnetic order down to 2 K can be observed. Thus, SmV$_6$Sn$_6$ can be viewed as a modest disordered derivative of the $R$V$_6$Sn$_6$ structure, in which the disordered rare earth ions can suppress the magnetic order and charge density wave in the $R$V$_6$Sn$_6$ kagome family.

\end{abstract}

\maketitle

\section{INTRODUCTION}

Kagome materials $RM_6X_6$ ($R$ = rare earth, $M$ = transition metal, $X$ = Sn, Ge) have drawn considerable attention in condensed matter physics due to their rich exotic magnetic and electronic properties\cite{peng2021,yin2020,ma2021,li2021,Jeonghun2022,venturini1992,venturini1991,teng2022,schobinger1997,szytula2004,rao1997} . These materials feature ideal kagome layer structures consisting of pure transition metals and naturally exhibit Dirac cone, saddle point, and flat bands that are characteristic electronic features of a kagome lattice\cite{li2021,gu2022,wang2020,peng2021,hu2022}. For a magnetic kagome system, such as $R$Mn$_6$Sn$_6$, the magnetic structure of Mn-sublattice can be flexibly modulated by the rare earth $R$ bearing different 4$f$ moments and ion anisotropy. When $R$ = Gd to Ho, ferrimagnetic (FIM) structures are formed, while antiferromagnetic (AFM) structures are formed when $R$ = Er, Tm, and Lu\cite{ma2021,wangbin2022}. In YMn$_6$Sn$_6$, frustrated interplanar exchange interactions were suggested to trigger strong magnetic fluctuations, resulting in a complex magnetic phase diagram\cite{dally2021,wang2021,ghimire2020}. Of particular interest, spin-orbit-coupled kagome lattice with strong out-of-plane magnetization can achieve spin-polarized Dirac dispersion with a large Chern gap in TbMn$_6$Sn$_6$\cite{yin2020}. The Chern-gapped Dirac fermions will host large Berry curvature and contribute to anomalous electronic and thermal trasport effects\cite{yin2020,xu2022,zhang2022E}. Besides, rare-earth magnetism can effectively engineer the Chern gap and related Hall conductivity\cite{ma2021}.

Compared with the robust magnetically ordered $R$Mn$_6$Sn$_6$, $R$V$_6$Sn$_6$ compounds solely exhibit weak magnetic couplings among the local 4$f$ moments of the $R$-sublattice. $R$V$_6$Sn$_6$ undergo magnetic ordering transitions at rather low temperatures.
The ordering temperature is below 6 K for $R$ = Gd-Ho, while no magnetic ordering is observed down to 1.8 K for $R$ = Er-Yb\cite{ishikawa2021,pokharel2021,rosenberg2022,zhang2022, Jeonghun2022,guo2022,Ganesh2022}. The crystalline electric field (CEF) of the rare earth ions is suggested to play an important role in the development of magnetic anisotropy\cite{zhang2022,Jeonghun2022}.
Topologically non-trivial Dirac surface state (GdV$_6$Sn$_6$ and HoV$_6$Sn$_6$)\cite{pokharel2021,peng2021,huyong2022}, quantum oscillation (YV$_6$Sn$_6$)\cite{pokharel2021}, quantum critical behavior (YbV$_6$Sn$_6$)\cite{guo2022}, and charge density wave (CDW) (ScV$_6$Sn$_6$)\cite{arachchige2022,zhan2022,hu2022} have been realized in the $R$V$_6$Sn$_6$ system.
These results suggest that the $R$ sublattice plays an important role in tuning the magnetic and electronic properties of $R$V$_6$Sn$_6$, driving us to explore the $R$-site substitution. $RM_6X_6$ with light $R$ may stablize differently crystal and magnetic structures. For example, SmMn$_6$Sn$_6$ and SmFe$_6$Sn$_6$ show the HoFe$_6$Sn$_6$ and YCo$_6$Ge$_6$-type structures, respectively\cite{stepien2000}. Furthermore, SmMn$_6$Sn$_6$ displays ferromagnetism and large intrinsic anomalous Hall effect\cite{MaWenlong2021}.

In this work, we report the crystal growth, structure, magnetism, electronic transport properties, and band structure calculations of a newly discovered kagome metal SmV$_6$Sn$_6$. Differing from the HfFe$_6$Ge$_6$-type structure of $R$V$_6$Sn$_6$ ($R$ = Gd-Lu, Y, Sc), SmV$_6$Sn$_6$ crystallizes in the hexagonal SmMn$_6$Sn$_6$ prototype\cite{malaman1997}. No evidence for magnetic order and CDW can be observed above 2 K, which may be ascribed to the structural disorders of Sm.
As a result of the CEF effect, the easy magnetization direction lies within the $ab$ plane.
Electronic transport measurements reveal multiband behaviors arising from the electronic bands of the vanadium kagome sublattice. Density functional theory (DFT) calculations reveal flat bands, Dirac cones, and saddle points for the  kagome V-3$d$ bands. The results highlight the interplay between the structural geometry, magnetism, and electronic correlations.

\section{EXPERIMENTAL AND CALCULATION DETAILS}

Single crystals of SmV$_6$Sn$_6$ were grown by the flux method\cite{pokharel2021}. The starting elements with a molar ratio of Sm : V : Sn $= 1:6:40$ were loaded in an alumina crucible and sealed in an evacuated silica ampoule. The ampoule was heated to 1100 $^\circ$C and held for 1 day, then cooled slowly at 3$^\circ$C/h to 650$^\circ$C. The ampoule was taken out and loaded into a centrifuge quickly to separate the excess Sn flux. A dilute HCl solution was used to remove the residual Sn from the crystal surface. The obtained single crystals have a typical size of $2\times1.5\times0.1$ mm$^3$ as shown in the inset of  Fig. \ref{fig1}(a).

Single-crystal x-ray diffraction (XRD) was conducted on a Bruker APEX-II CCD x-ray diffractometer at 150 K. Powder XRD pattern of a crushed single crystal was collected on a powder XRD diffractometer (Empyrean) at room temperature. The elemental analysis was carried out using an energy dispersive x-ray spectroscopy (EDX) (EVO, Zeiss). Magnetic and electronic transport properties were measured on a physical property measurement system (Quantum Design). The in-plane resistivity was measured using the standard four-probe method.

DFT calculations for the structural relaxation and further electronic property of SmV$_6$Sn$_6$ were performed using the Vienna ${Ab}$ ${initio}$ Simulation Package (vasp5.4.4)\cite{kresse1996}. The exchange and correlation function was described using the Perdew-Burke-Ernzerhof generalized gradient approximation\cite{perdew1981}. The all-electron projector augmented wave method\cite{kresse1999} was adopted to describe the electron-ion interactions, treating 5$s$$^2$5$p$$^2$, 3$d$$^3$4$s$$^2$ and 5$s$$^2$5$p$$^6$4$f$$^5$5$d$$^1$6$s$$^2$ as valence electrons for Sn, V and Sm atoms, respectively. To ensure convergence of the total energy within a precision of 1 meV per atom, the plane-wave energy cutoff was set at 500 eV, and the Monkhorst-Pack\cite{monkhorst1976} $k$-point grids with a reciprocal space resolution of 2$\pi$ $\times$ 0.03 {\AA}$^{-1}$ in the Brillouin zone were selected. Structural relaxations were performed with forces converged to less than 0.001 eV{\AA}$^{-1}$.

\begin{figure}[t]
\includegraphics[scale = 0.45]{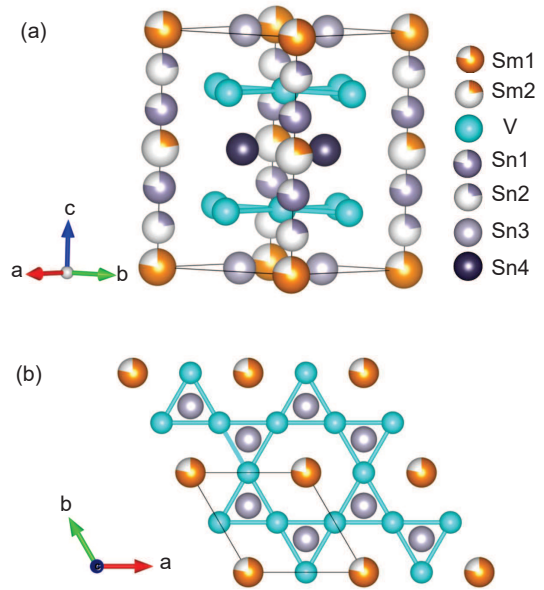}
\caption{ (a) Three-dimensional crystal structure of SmV$_6$Sn$_6$. The atom balls with different colors suggest that atoms have fractional occupancy. (b) Top view of the crystal structure, showing the kagome lattice consisting of V atoms. }
\label{fig2}
\end{figure}

\begin{figure}[t]
\includegraphics[scale = 0.45]{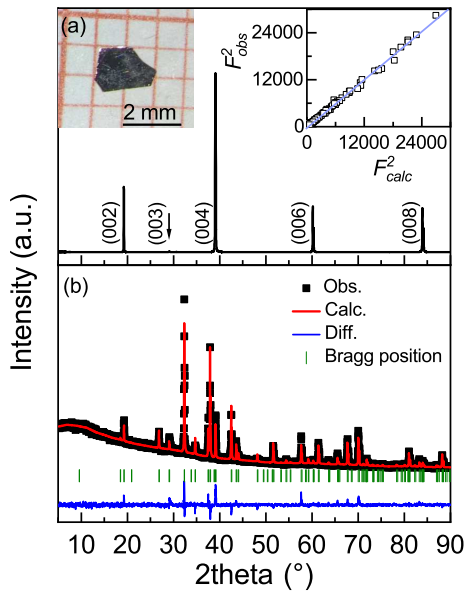}
\caption{ (a) XRD measurement on the $ab$ plane of a single crystal of SmV$_6$Sn$_6$. The inset on the left is a picture of the single crystal and the right inset shows the refinement of single-crystal diffraction. (b) A pattern of XRD measured on powder samples. }
\label{fig1}
\end{figure}

\section{RESULTS AND DISCUSSIONS}
\subsection{Crystal structure}

Single-crystal XRD measurements reveal SmV$_6$Sn$_6$ crystallizes in the hexagonal space group $P$6/$mmm$ as that of the SmMn$_6$Sn$_6$\cite{malaman1997}, as shown in Fig. \ref{fig2}. The ideal V-kagome layers are separated by Sm and Sn atoms.  The refined structural parameters were summarized in Table \ref{table:t1}. The measured structure factors are in good agreement with the calculated structure factors as shown in the inset of Fig. \ref{fig1}(a).

The results yield 22\% of the Sm ions occupying on the intermediate lattice positions (1$b$ site) that are shifted from the Sm positions (1$a$) by $c$/2 along the $c$ axis. The Sm ions on the intermediate sites enforce the nearest neighbor Sn ions to the Sn2 sites. The occupancy of Sm in both the Sm1 (1$a$ site) and Sm2 (1$b$ site) positions is confirmed by the weaker intensity of the (003) reflection compared with the fully ordered GdV$_6$Sn$_6$ and HoV$_6$Sn$_6$, as shown in Fig. \ref{fig1}(a)\cite{peng2021}. The shift of the $R$ atoms has also been observed in the other $RM_6$Sn$_6$ systems, where $R$ are light rare earth elements\cite{malaman1997,stepien2000}.  Following the lanthanide contraction rule, the lattice constants $a$ and $c$ in SmV$_6$Sn$_6$ are larger than that of the heavy rare earth elements, indicating the Sm ions are trivalent\cite{Jeonghun2022}. Thus,  SmV$_6$Sn$_6$ can be regarded as a modest disordered derivative of the $R$V$_6$Sn$_6$ ($R$ = Gd-Lu, Y, Sc) structure. The powder XRD pattern in Fig. \ref{fig1}(b) further confirms that no impurity could be identified within our instrumental resolution. Besides, the EDX measurements yield an atomic ratio of Sm: V: Sn $= 1: 6.79: 6.62$, close to the stoichiometry of SmV$_6$Sn$_6$.

\begin{table}[t]
\caption{Crystal parameters of SmV$_6$Sn$_6$ from the single-crystal XRD at 150 K (Ga K$\alpha$, $\lambda$ = 1.34138 {\AA})}
\begin{tabular}{ccccccc}
\hline \hline
Formula            & SmV$_6$Sn$_6$   & occupancy    \\ \hline
Space group   & $P$6/$mmm$      &        \\
a/{\AA}   & 5. 5243(6)      &         \\
c/{\AA}   & 9.188(2)      &         \\
Sm1 & 1$a$(0, 0, 0)      & 0.78        \\
Sm2 & 1$b$(0, 0, 1/2)      & 0.22        \\
V & 6$i$(0, 0, 1/4)      & 1        \\
Sn1 & 2$e$(0, 0, 0.3356(2))      & 0.78        \\
Sn2 & 2$e$(0, 0, 0.1673(9))      & 0.22        \\
Sn3 & 2$c$(1/3, 2/3, 0)      & 1        \\
Sn4 & 2$d$(1/3, 2/3, 1/2)      & 1        \\
Goodness-of-fit on $F$$^2$ & 1.112      &          \\
$R$$_1$, w$R$$_2$ (all data) & 0.0378, 0.0747      &          \\ \hline \hline
\end{tabular}
\label{table:t1}
\end{table}

\subsection{Magnetic properties}
\begin{figure}[t]
\includegraphics[scale = 0.36]{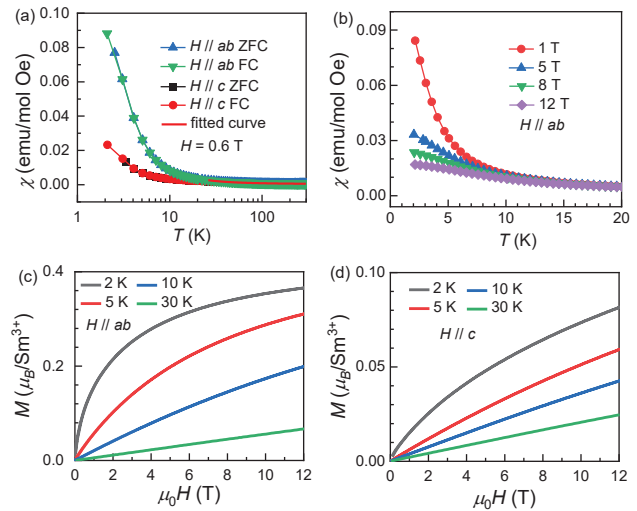}
\caption{ (a) Temperature dependence of magnetic susceptibilities measured with a 0.6 T magnetic field along the $c$ axis and $ab$ plane of SmV$_6$Sn$_6$. (b) Magnetic susceptibilities measured with various magnetic fields along the $ab$ plane. (c) Magnetization against magnetic fields along the $c$ axis and (d) the $ab$ plane at 2, 5, 10, and 30 K. }
\label{fig3}
\end{figure}

Figures \ref{fig3}(a) and \ref{fig3}(b) show magnetic susceptibility [$\chi$($T$)] measured with a 0.6 T magnetic field. No obvious furcation exists between zero-field-cooling (ZFC) and field-cooling (FC) susceptibility curves. The $\chi$($T$) along the $ab$-plane is larger than that along the $c$-axis at low temperatures, revealing a strong easy-plane anisotropy.
The FC susceptibility ($25\le T\le300$ K) follows the expression:
 $\chi$ = $C$/($T$ $-$ $\theta$$_{CW}$) $+$ $\chi$$_{VV}$, where $\theta$$_{CW}$ is the Curie-Weiss temperature, $C$ is the Curie constant, and $\chi$$_{VV}$ is the temperature-independent Van Vleck contribution\cite{hodges2001}.
 Our fitting yields $\theta$$_{CW}^{ab}$ = 8.1(5) K, $C$$^{ab}$ = 3.99(9) $\times$ 10$^{-2}$ emu K mol$^{-1}$ Oe$^{-1}$, and $\chi$$_{VV}$ = 4.52(7) $\times$ 10$^{-4}$ emu mol$^{-1}$ Oe$^{-1}$ for $H$ $\|$ $ab$ plane; $\theta$$_{CW}^c$ = $-$6.1(8) K, $C$$^c$ = 3.86(9) $\times$ 10$^{-2}$ emu K mol$^{-1}$ Oe$^{-1}$, and $\chi$$_{VV}$ = 7.56(5) $\times$ 10$^{-4}$ emu mol$^{-1}$ Oe$^{-1}$ for $H$ $\|$ $c$ axis. The positive $\theta$$_{CW}^{ab}$ suggests dominant FM-type in-plane magnetic interaction while the negative $\theta$$_{CW}^c$ shows AFM-type out-of-plane interactions. The effective magnetic moments of Sm$^{3+}$ ions are calculated to be $\mu$$_{eff}^{ab}$ = 0.565(1) $\mu$$_B$ and $\mu$$_{eff}^c$ = 0.556(1) $\mu$$_B$, which are smaller than 0.845 $\mu$$_B$ of a free Sm$^{3+}$ ion ($J$ = 5/2, $g$ = 2/7). The reduced effective moments are also observed in the other samarium compounds such as Sm$_3$Sb$_3$Zn$_2$O$_{14}$ (0.53 $\mu$$_B$)\cite{Sanders2008} and Sm$_2$Zr$_2$O$_7$ (0.5 $\mu$$_B$)\cite{singh2008}, which are likely related to the narrow $J$ multiplet width arising from the large crystal field splitting of the lowest $J$ = 5/2 multiplets. The magnetic susceptibility increases rapidly and large anisotropy is developed below 10 K, consistent with the development of short-range magnetic order. No evidence for a magnetic ordering transition appears down to 2 K from susceptibility measured under in-plane magnetic fields, as shown in Fig. \ref{fig3}(b).

Isothermal magnetization data at different temperatures are plotted in Figs. \ref{fig3}(c) and \ref{fig3}(d). Large magnetic anisotropy can be observed at low temperatures. The magnetization for $H$ $\|$ $ab$ plane is about 4.5 times larger than that for $H$ $\|$ $c$ at 2 K. When the field parallel to the $ab$ plane at 2 K, the magnetization increases linearly in the low field and then gradually approach saturation. The magnetization is up to 0.37 $\mu$$_B$ at 12 T, which is far smaller than the theoretical value of $M$$_s$ = $gJ$ = 0.71 $\mu$$_B$. The magnetization increases linearly for temperatures above 10 K, suggesting that the short-range order disappears at high temperatures. The magnetic anisotropy at low temperatures could be ascribed to the CEF.

\begin{figure}[t]
\includegraphics[scale = 0.3]{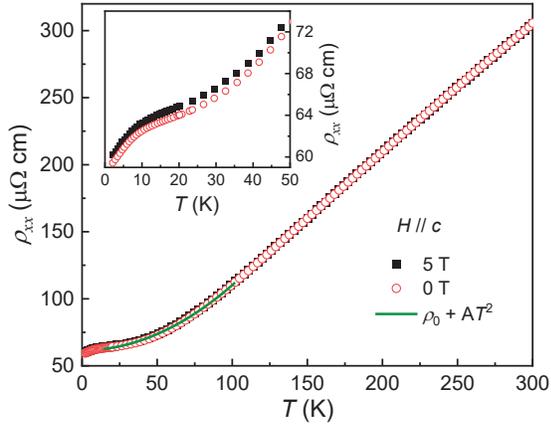}
\caption{ Resistivity of SmV$_6$Sn$_6$ measured at 0 and 5 T magnetic fields. The green curve shows the fitting of ($\rho$$_{xx}$(T)) via the equation $\rho$ = $\rho$$_0$ + ${AT}$$^2$. The inset is a zoom-in at low temperatures. }
\label{fig4}
\end{figure}

\subsection{Transport properties}
Figure \ref{fig4} shows the temperature dependence of the in-plane resistivity $\rho$$_{xx}$($T$) at 0 and 5 T magnetic fields. The resistivity decreases with decreasing temperature, consistent with the characteristic of a metal. The resistivity in the temperature range of 10$-$100 K follows the Fermi liquid behavior $\rho$ = $\rho$$_0$ + ${AT}$$^2$, giving rise to $\rho$$_0$ = 61.58(1) $\mu\Omega$ cm and $A$ = 4.86(1)$\times$10$^{-3}$ $\mu\Omega$ cm K$^{-2}$. The inset of Fig. \ref{fig4} displays a decrease of resistivity below 10 K, which may be related to the formation of short-range magnetic order. In addition, the resistivity is enhanced by a 5 T magnetic field at low temperatures.

\begin{figure}[t]
\includegraphics[scale = 0.35]{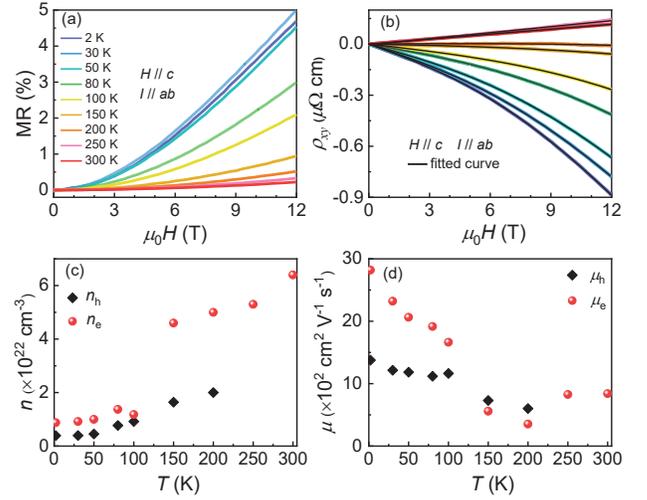}
\caption{ (a) Magnetoresistance at various temperatures measured with $H$ $\|$ $c$ axis. The magnetoresistance component was isolated by averaging $\rho$$_{xx}$($H$) and $\rho$$_{xx}$($-H$). (b) Hall resistivity $\rho$$_x$$_y$ measured at different temperatures. To remove the effects of the longitudinal resistivity, $\rho$$_x$$_y$ is obtained via $\rho$$_x$$_y$ = [$\rho$$_x$$_y$($H$)$-$$\rho$$_x$$_y$($-$$H$)]/2. (c) Carrier densities of electrons and holes obtained from fitting of the Hall resistivity in (b) with a two-band model below 250 K and a single-band model for higher temperatures. (d) Charge carrier mobilities aganist temperature obtained from fitting the Hall resistivity.}
\label{fig5}
\end{figure}

To study the magnetoresistance (MR) effect, we further measured field dependence of the longitudinal resistivity [$\rho$$_{xx}$($H$)] at various temperatures. The MR defined as [$\rho_{xx}(H)-\rho_{xx}(H=0)]/\rho_{xx}(H=0)\times 100$\% is shown in Fig. \ref{fig5}(a). The current is along the $ab$ plane and the magnetic field is along the $c$ axis. A small magnitude of MR can be observed at all temperatures. The MR does not tend to saturate up to 12 T. The MR is gradually increased with temperature decreasing. However, a small decrease can be found at 2 K compared with that of 30 K, which may be a consequence of suppressed spin-disorder scattering.

To characterize the carrier concentrations and effective mobilities, Hall resistance [$\rho$$_{xy}$($H$)] measurements were performed from 2 to 300 K with the magnetic field applied parallel to the $c$-axis and the current within the $ab$ plane as shown in Fig. \ref{fig5}(b).  The $\rho$$_{xy}$($H$) curves are linear at temperatures $T\geq250$ K. While the $\rho$$_{xy}$ decreases nonlinearly as the tempeature decreases below 250 K, signifying a multiband transport behavior. The $\rho$$_{xy}$($H$) curves can be fitted with a single-band model at high temperatures ($T\geq$ 250 K) and a two-band model at low temperatures ($T<250$ K). Figures \ref{fig5}(c) and \ref{fig5}(d) display the fitted results. It can be found that the concentration and mobility of the electron carriers are larger than that of the holes for most temperatures, suggesting that electrons play a dominant role in conduction. The similar multiband transport behaviors have also been observed in $R$V$_6$Sn$_6$ ($R$ = Gd-Lu, Y)\cite{pokharel2021,Ganesh2022,zhang2022}.

\begin{figure*}[t]
\centering
\includegraphics[scale = 0.5]{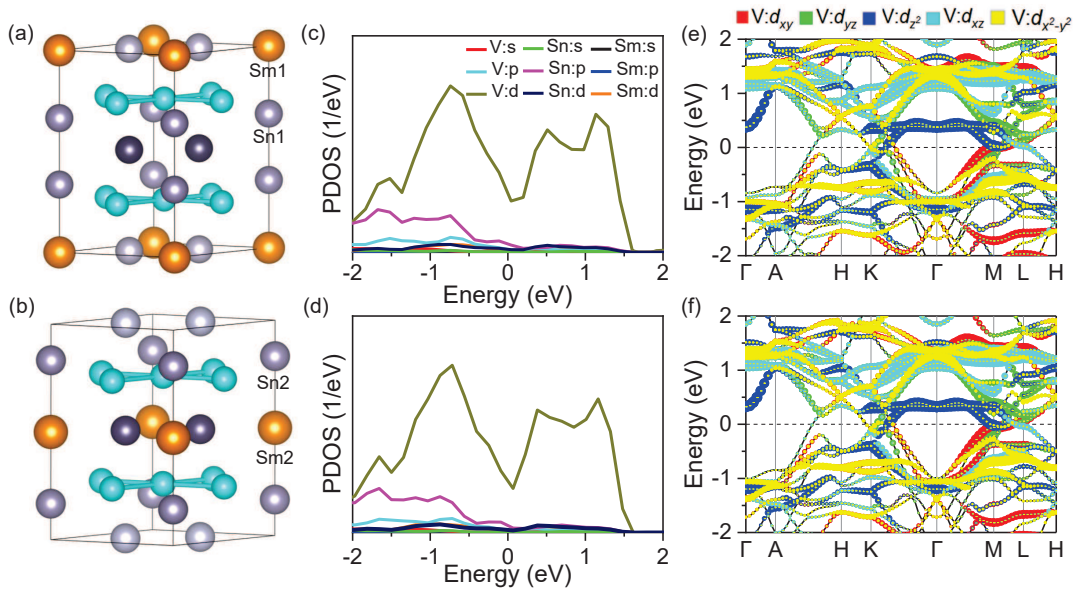}
\caption{ (a) Sketch of the type-1 structure with fully occupied Sm1 and Sn1 sites. (b) Sketch of the type-2 structure with Sm and Sn occupying on the Sm2 and Sn2 sites. (c) Orbital projected density of states of type-1 and (d) type-2 structures. (e) and (f) V $d$ orbital decomposed band structures. }

\label{fig6}
\end{figure*}

\subsection{Electronic structure}
It is difficult to calculate the electronic structure via the DFT for the compound with disorders. We constructed two types of crystal structures via rearranging the atomic occupancies of the Sm and Sn sites in SmV$_6$Sn$_6$. For the type-1 structure, the atomic occupancies of both the Sm1 and Sn1 sites are 100$\%$; for the type-2 structure, Sm and Sn fully occupy on the Sm2 and Sn2 sites, respectively. The experimentally synthesized SmV$_6$Sn$_6$ can be regarded as an intermediate phase of the type-1 and type-2 structures. Figures \ref{fig6}(c)-\ref{fig6}(f) show that the two types of structures possess similarly electronic structures, suggesting the synthesized SmV$_6$Sn$_6$ has an approximative electronic structure. The orbital projected density of states yields that the 3$d$ electrons of V atoms make a dominant contribution near the Fermi level. The electronic band structures are shown in Figs. \ref{fig6}(e) and \ref{fig6}(f), an obvious kagome flat band characterized by the V $d$$_{z^2}$ orbitals can be observed above $E_F$. Moreover, the Dirac cones and saddle points appear at the $K$ and $M$ points, respectively, as expected from minimal kagome tight-binding models\cite{yin2022}. Such the transition metal-based $d$-orbital band structure endemic to the kagome lattice in the proximity of the Fermi level is one of the common features in the family of $R$$M$$_6$Sn$_6$\cite{li2021,gu2022,wang2020,peng2021,hu2022}.

\section{DISCUSSIONS}

Compared with the $R$$M$$_6$$X$$_6$ ($M$ = Mn, Fe, $X$ = Ge, Sn) family with magnetic ordering temperatures originated from the transition metals well above the room temperature\cite{venturini1992,venturini1991,rao1997}, the moments of the rare earth ions in $R$V$_6$Sn$_6$ ($R$ = Gd-Ho) order at rather low temperatures \cite{Jeonghun2022}. The magnetic transition temperatures of $R$V$_6$Sn$_6$ are affected by the CEF of the $R$ ions. The leading crystal field parameter $B_2^0$ can be calculated via the point charge model $B_2^0$ = 10($\theta$$_{CW}^{ab}$$-$$\theta$$_{CW}^c$)/3(2$J$$-$1)(2$J$+3), giving rise to $B$$_{2}^{0}$ = 1.48 K for SmV$_6$Sn$_6$. The positive value is close to that of TmV$_6$Sn$_6$\cite{Jeonghun2022}, which shows an easy in-plane anisotropy. Normally, a large $B_2^0$ with strong CEF as like -1.4 K of TbV$_6$Sn$_6$ is expected to enhance the magnetic ordering temperature\cite{Jeonghun2022}. However, the magnetic order has not been observed down to 2 K in SmV$_6$Sn$_6$. The absence of magnetic order may be related to the atomic disorder associated with the shifts of the Sm$^{3+}$ ions along the $c$ axis.

The DFT calculations reveal that the flat bands at $\Gamma$ point of the type-1 and type-2 structures locate at 0.376 and  0.326 eV, respectively. The difference should be associated with the strength of V $d-$Sn $p$ orbital hybridization.
The different V-Sn1 (2.872(1) \AA) and V-Sn2 (2.865(3) \AA) bond lengths will result in different strength of orbital hybridization. It was reported that ScV$_6$Sn$_6$ undergoes a CDW phase transition at 92 K, in which the V-Sn bond length is 2.822(1) \AA\cite{arachchige2022}. While no evidence of CDW is observed in the isostructural $R$V$_6$Sn$_6$ compounds with the larger ions $R$ = Y, Sm$-$Lu\cite{Jeonghun2022,guo2022}. The V-Sn bond length in the $R$V$_6$Sn$_6$ family depends on the chemical pressure arising from the rare earth ions $R$$^{3+}$. The smaller the ion radius of $R$$^{3+}$, the shorter the bond length. Thus, further explorations for smaller $R$ systems or the effect of pressure are interesting for investigatations of the interplays among lattice, charge, and magnetism.

\section{CONCLUSIONS}
The single crystals of the kagome metal SmV$_6$Sn$_6$ have been grown by the self-flux method. The crystal structure of SmV$_6$Sn$_6$ belongs to the SmMn$_6$Sn$_6$ prototype, showing partial crystallographic disorder associated with the appearance of the additional Sm site. Large magnetic anisotropy due to the CEF effects is found for the magnetic Sm$^{3+}$ ions. Moments tend to align in the $ab$ plane. No abnormal transition related to long-range magnetic ordering and CDW can be observed down to 2 K. Electronic transport measurements exhibit multiband behaviors and positive magnetoresistance properties. DFT calculations show a V-3$d$ kagome band structure near the Fermi level with a flat band, Dirac cones, and saddle points.

\section{ACKNOWLEDGMENTS}

Work was supported by the Guangdong Basic and Applied Basic Research Foundation (Grants No. 2021B1515120015, 2022A1515010035), the National Natural Science Foundation of China (Grants No. 12174454, U213010013), the Guangzhou Basic and Applied Basic Research Foundation (Grants No. 202201011123, 202201011798), Guangdong Provincial Key Laboratory of Magnetoelectric Physics and Devices (Grant No. 2022B1212010008), and National Key Research and Development Program of China (Grant No. 2019YFA0705702).

\bibliography{ref}
\end{document}